\documentclass[twocolumn,showpacs,superscriptaddress,nofootinbib,aps,prb,floatfix]{revtex4-1}
\usepackage{graphicx}
\usepackage{amsmath,amssymb}
\usepackage[utf8]{inputenc}
\usepackage{natbib}
\usepackage{hyperref}
\usepackage{color}
\hypersetup{colorlinks=true,citecolor={blue},linkcolor={blue},urlcolor={blue}}
\usepackage{units}
\usepackage[english]{babel}
\usepackage{upgreek}
\usepackage{dcolumn}
\usepackage{bm}
\usepackage{xfrac}

\usepackage[normalem]{ulem}
\usepackage{calrsfs}
\usepackage{nicefrac}

\usepackage{soul,xcolor}

\setstcolor{red}

\begin{document}
\renewcommand{\thefootnote}{\alph{footnote}}	
\title{Optical control of synchronous phases in a programmable polariton cell}
\date{\today}
\author{S. Alyatkin}
\affiliation{Skolkovo Institute of Science and Technology, Novaya St. 100, Skolkovo 143025, Russian Federation}
\author{J. D. T\"opfer}
\affiliation{Skolkovo Institute of Science and Technology, Novaya St. 100, Skolkovo 143025, Russian Federation}
\affiliation{School of Physics and Astronomy, University of Southampton, Southampton, SO17 1BJ, United Kingdom}
\author{A. Askitopoulos}
\affiliation{Skolkovo Institute of Science and Technology, Novaya St. 100, Skolkovo 143025, Russian Federation}
\author{H. Sigurdsson}
\affiliation{Skolkovo Institute of Science and Technology, Novaya St. 100, Skolkovo 143025, Russian Federation}
\affiliation{School of Physics and Astronomy, University of Southampton, Southampton, SO17 1BJ, United Kingdom}
\author{P. G. Lagoudakis}
\email{Pavlos.Lagoudakis@soton.ac.uk}
\affiliation{Skolkovo Institute of Science and Technology, Novaya St. 100, Skolkovo 143025, Russian Federation}
\affiliation{School of Physics and Astronomy, University of Southampton, Southampton, SO17 1BJ, United Kingdom}

\begin{abstract}
We demonstrate deterministic control of the nearest and next-nearest neighbor coupling in the unit cell of a square lattice of microcavity exciton-polariton condensates. We tune the coupling in a continuous and reversible manner by optically imprinting potential barriers of variable height, in the form of spatially localized incoherent exciton reservoirs that modify the particle flow between condensates. By controlling the couplings in a $2\times2$ polariton cluster, we realize ferromagnetic, anti-ferromagnetic and paired ferromagnetic phases. Our approach paves the way towards simulating complex condensed matter phases through precise control of the individual couplings in networks of optical nonlinear oscillators.

\end{abstract}


\maketitle
\section{Introduction}
Control over a novel type of a many-body optical network, the exciton-polariton condensate lattice, is highly desirable partly due to its  application in integrated optical circuitry~\cite{Sanvitto2016}, potential for quantum computation~\cite{Byrnes_PRA2012, Cuevas_Science2018} as well as a testbed for the study of emergent phenomena in large complex systems; these range from the Kibble-Zurek mechanism~\cite{Solnyshkov_PRL2016}, spontaneous magnetization~\cite{ohadi_spin_2017}, topological phases~\cite{klembt_exciton-polariton_2018}, to reverse ground state annealing~\cite{berloff_realizing_2017}. Indeed, exciton-polaritons (from here on {\it polaritons}) have already found a role in various memory processing elements such as logic gates~\cite{Leyder_PRL2007}, transistors~\cite{gao_polariton_2012, zasedatelev_room-temperature_2019}, switches~\cite{Amo_NatPho2010}, routers~\cite{Marsault_APL2015}, and diodes~\cite{Nguyen_PRL2013}, ranging from cryogenic to room-temperature operation conditions. The accuracy with which one can deterministically control the interactions between individual polariton condensates reflects the experimental control with which one can design advanced polaritonic devices, and access interesting phases of interacting many-body systems.

Polariton condensates~\cite{Carusotto_RMP2013} are described as coherent ensembles of bosonic light-matter quasi-particles that in the mean field treatment, are described by a coherent macroscopic wavefunction dictated by a complex nonlinear Schrödinger equation. Interactions between trapped polariton condensates can be accurately quantified using the tight-binding treatment, where the condensates can display synchronization~\cite{ohadi_synchronization_2018}, Josephson oscillations~\cite{Abbarchi_NatPhy2013}, frequency combs~\cite{Rayanov_PRL2015}, and other intriguing effects stemming from their non-Hermitian and strong nonlinear character. Relevant to the current study, the interaction strength between spatially separated polariton condensates can be dramatically enhanced by realizing them not in a trapping geometry but instead as stable ballistically expanding optical fluids sustained by the gain of the excitation source~\cite{Wouters_2008PRB, topfer_towards_2019,ohadi_nontrivial_2016}. Recently, a study on the interactions between expanding polariton condensates revealed that their dynamics are dictated by time-delay equations of motion similar to the Lang-Kobayashi equation  used to describe coupled laser systems~\cite{Kobayashi_JQE1980, topfer_towards_2019}. Therefore, designing a network of these expanding polariton condensates brings in a new platform for the study of nonlinear oscillatory systems used to describe chaos, neurological functions, social behavior, and synchronization~\cite{pikovsky2003synchronization}. On the other hand, inspired by recent developments on Ising machines based on optical parametric oscillators~\cite{marandi_network_2014,mcmahon_fully_2016}, networks of polariton condensates can potentially be designed to tackle computationally challenging problems in an analogue manner by associating the degrees of freedom of the condensates (i.e., amplitude, phase, and polarization) to appropriate spin Hamiltonians~\cite{berloff_realizing_2017} that can be mapped to computationally complex tasks~\cite{barahona_computational_1982}. Either case would require programmable coupling strengths between condensates. Recently, gates with dissipation in-between condensate nodes were proposed for arbitrary interaction control in condensate lattices~\cite{kalinin2019towards}.

In this paper, we demonstrate an all-optical method to tune and measure the coupling between adjacent polariton condensates. Different from lattices of polariton condensates realized in lithographically written structures~\cite{schneider_exciton-polariton_2016}, here we optically imprint networks of polariton condensates using tightly focused non-resonant optical beams in a planar microcavity shaped by a spatial light modulator (SLM). This allows for both precise and rewritable control of the excitation profile that makes it possible to drive condensate networks of various geometries in a microcavity. Furthermore, we additionally imprint optically incoherent exciton reservoirs of controllable density along the edges connecting polariton condensates. The repulsive interaction between reservoir excitons and condensate polaritons acts as a barrier that modifies the polariton flow between adjacent condensates~\cite{amo2010light,gao_polariton_2012,askitopoulos_polariton_2013}. We show that by changing the excitation power of these excitonic barriers, and subsequently their potential height, we can tune individual couplings between polariton condensates. We implement this concept to demonstrate a programmable $2\times2$ polariton condensate unit cell.

\section{Results}
\begin{figure}[t!]
	\includegraphics[scale=1]{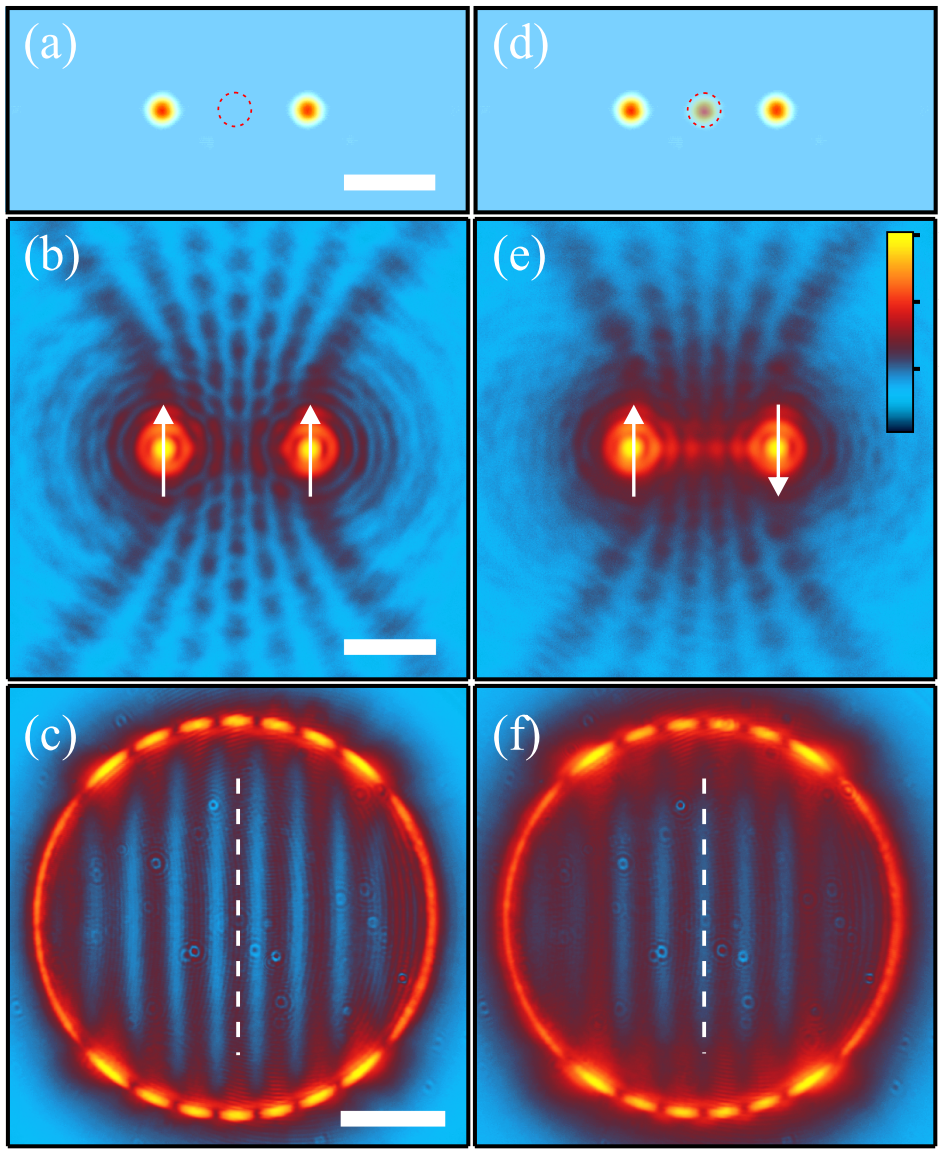}
	\caption{(a) Intensity profile of two non-resonant excitation spots used to imprint a polariton dyad in a FM configuration with corresponding normalized PL intensity shown in both (b) real- and (c) reciprocal-space in a logarithmic color scale. Introducing an optically-imprinted weak barrier in between the two condensates (d) switches the phase configuration of the dyad to an AFM state as shown (e,f) by the change in the interference pattern. White scale bars correspond to $10\;\mathrm{\upmu m}$, and $1\;\mathrm{\upmu m}^{-1}$ in real and reciprocal space respectively. Red dashed circles in (a,d) denote barrier position. White dashed vertical lines in (c,f) are guide to the eye.}
	\label{fig1}
\end{figure}

We use a strain compensated 2$\lambda$ GaAs based planar microcavity with embedded three pairs of In$_{0.08}$Ga$_{0.92}$As quantum wells described in Ref.~[\onlinecite{cilibrizzi_polariton_2014}]. The sample is held at $\approx 4\;\mathrm{K}$ within a closed-cycle helium cryostat. The excitation profile is controlled with a reflective phase-only SLM which modulates the incident laser beam. We start with the simplest building block of the condensate lattice -- a {\it polariton dyad}. Figure~\ref{fig1}(a) shows the non-resonant ($\lambda = 796\; \mathrm{nm}$) pump profile consisting of two laser spots that are tightly focused (FWHM$\approx 1.6\;\mathrm{\upmu m}$) with a microscope objective of NA=0.42 and separated by a distance $d\approx 15.7\;\mathrm{\upmu m}$. Each of the two condensates is driven at a power of $1.3\times \mathrm{P}_{\mathrm{thr}}$, where $\mathrm{P}_{\mathrm{thr}}$ is the threshold power determined for a single isolated condensate. The constructive interference in the real space photoluminescence (PL) at the center between the two condensates [Fig.~\ref{fig1}(b)], and the bright central vertical fringe in the reciprocal space PL [white dashed line in Fig.~\ref{fig1}(c)] indicate in-phase locking of the dyad \cite{ohadi_nontrivial_2016}. Here, we denote in-phase and anti-phase configuration with parallel and anti-parallel white arrows (spins) resembling ferromagnetic (FM) and anti-ferromagnetic (AFM) type of arrangement of phases between the condensates.

Next, we investigate the effect of an optically induced barrier on the coupling of the dyad. A third laser beam injects non-resonantly an exciton reservoir in the middle of the dyad. We use cross-circularly polarized excitation for the barrier with respect to the condensates pumps in order to minimize gain due to overlap of the condensate wavefunctions with the barrier. The cross-circular configuration allows for a wider tunability of the barrier exciton density before barrier-induced nonlinearities set in. Figure~\ref{fig1}(d) shows the pumping profile for a polariton dyad in the presence of a weak barrier pumped below threshold with $\mathrm{P}_\mathrm{bar}=0.36\times \mathrm{P}_{\mathrm{thr}}$. The resulting interference patterns of the polariton PL in real space [Fig.~\ref{fig1}(e)] and reciprocal space [Fig.~\ref{fig1}(f)] reveal that the phase configuration in the presence of the barrier has switched from FM to AFM. Numerical simulations producing the switching of parities are shown in Section~\ref{sec2pumps}.
\begin{figure*}[t!]
	\includegraphics[scale=1]{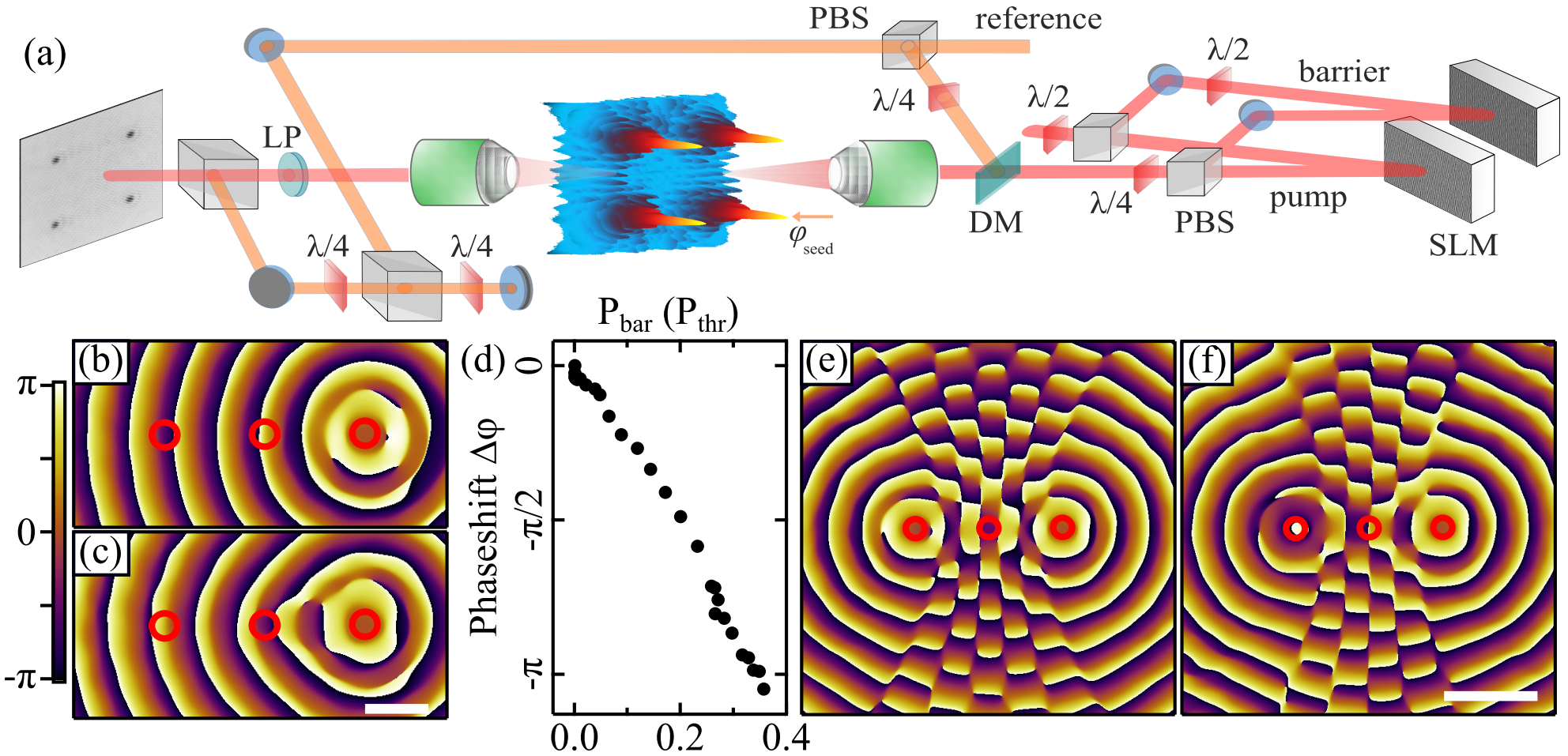}
	\caption{(a) Scheme of the setup that allows for simultaneous imprinting of polariton condensates and barriers giving all-optical control of couplings. An additional weak laser beam (\emph{seed}) resonant to the PL is injected to one condensate fixing its phase and is subsequently used as a reference wave for interferometric homodyne measurements to implement live phase readout. Measured phase maps of a single condensate (right-most red circle) pumped at $P=1.3\times P_{\mathrm{thr}}$ subject to a weak barrier (central red circle) with (b) $\mathrm{P}_{\mathrm{bar}} = 0$ and (c) $\mathrm{P}_{\mathrm{bar}} = 0.36\times \mathrm{P}_{\mathrm{thr}}$. (d) The measured phase shift $\Delta \phi$ integrated within a circle of $1\; \mathrm{\upmu m}$ radius (left-most red-circle) at a position $d \approx 15.7 \;\mathrm{\upmu m}$ away from the condensate shows a continuous decrease as a function of barrier pump power. The extracted phase maps (e,f) for the two configurations in Fig.~\ref{fig1} confirm the observed FM and AFM configurations by $0$ and $\pi$ phase differences between the two condensate centres. The white scale bar in (c) [applying also to (b)] amounts for $5\;\mathrm{\upmu m}$, while the scale bar in (f) [applying also to (e)] amounts for $10\;\mathrm{\upmu m}$.}
	\label{fig2}
\end{figure*}

The phase difference between two coupled ballistically expanding condensates can be extracted from the resulting interference patterns in both real- and reciprocal-space. However, in the case of more complex geometries and couplings that may lead to frustration in the system, phase retrieval from intensity maps becomes less well defined, while at the same time, accuracy in the phase retrieval is important for the application of polariton lattices as simulators. In the following, we develop a homodyne interferometric technique that utilizes the U(1) symmetry in the classical phase configuration of coupled polariton condensates, that allows for single-shot phase retrieval of all the condensates across a lattice.

The schematic in Fig.~\ref{fig2}(a) shows the experimental configuration of the homodyne technique. An additional weak resonant excitation beam is used to fix the phase of one condensate. The linewidth of the resonant excitation ($\approx$ 100 kHz) is much narrower than that of the polariton condensate. The emission of the whole polariton network is interfered with the resonant seed beam using a Mach-Zehnder interferometer. This allows to extract the full phase map of the lattice with off-axis digital holography \cite{kreis_digital_1986,liebling_complex-wave_2004}. We apply this technique both to the polariton dyad (Fig.~\ref{fig1}) and to the case of a single condensate, i.e. when one of the excitation pump spots of the dyad is blocked. The effect of the barrier onto the radial outflow of polaritons from a single condensate can be seen in Fig.~\ref{fig2} for (b) $\mathrm{P}_\mathrm{bar} = 0$ and (c) $\mathrm{P}_\mathrm{bar} = 0.36\times \mathrm{P}_\mathrm{thr}$. The red circles indicate, from right to left, the positions of a single pumped condensate, the barrier, and the mirrored location of the condensate with respect to the barrier. We subtract the obtained phase maps with and without the barrier, and extract the phase shift $\Delta \phi$ of the condensate wavefunction at the location of the left most circle as a function of the barrier pump power. Figure~\ref{fig2}(d) shows a continuous shift in $\Delta \phi$ up to $-\pi$ up to the same barrier pump power that switches the parity of the dyad (Fig.~\ref{fig1}). Figure \ref{fig2}(e,f) show the obtained phase maps corresponding to the pumping profiles of Fig.~\ref{fig1}(a,d). Numerical simulations producing the phase shift $\Delta \phi$ are shown in Section~\ref{sec1pump}.

\begin{figure}[t!]
	\includegraphics[scale=1]{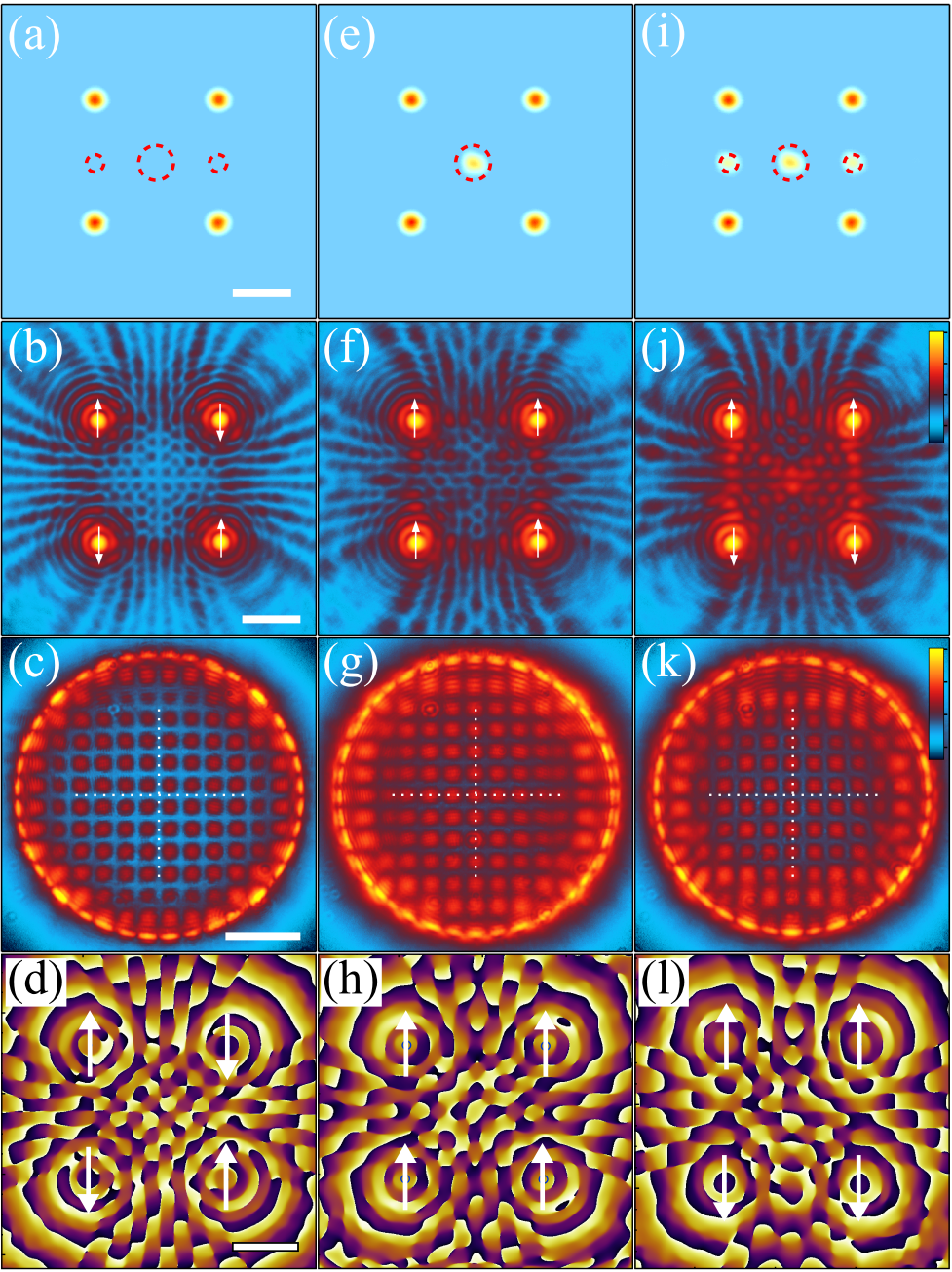}
		\caption{(a,e,i) Real-space intensity map of the non-resonant excitation geometry with barrier beams enclosed by red dashed circles. (b,f,j) Real- and (c,g,k) reciprocal-space PL of the polariton condensates generated by the different barrier configurations (a,e,i) respectively showing AFM, FM, and PFM phases. Corresponding real-space phase maps are given in (d,h,l) with white arrows denoting the magnetic arrangement of the cluster. The color scale in (d,h,l) is the same as in Fig.~\ref{fig2}.}
	\label{fig3}
\end{figure}

From the simplest configuration of the polariton dyad, we expand to a more complex system, the unit cell of a square lattice, wherein the coupling between condensates occurs not only with the nearest neighbors, but also with next nearest neighbors. Fig.~\ref{fig3}(a) shows the pumping profile of the $2\times2$ polariton cell. The pump beams are co-circularly polarized and fixed at $1.34 \times \mathrm{P}_\mathrm{thr}$. The distribution of the polariton real-space PL [Fig.~\ref{fig3}(b)] reveals AFM coupling between the four nodes, also corroborated by the reciprocal-space PL [Fig.~\ref{fig3}(c)]. Here, the lattice constant is chosen such that polariton condensation occurs at a single energy state.

We introduce a cross-circulary polarized pump in the center of the square that is kept at $0.5\times \mathrm{P}_{\mathrm{thr}}$ power [Fig.~\ref{fig3}(e)]. From the interference pattern in both real- and reciprocal-space PL we detect a switch from an AFM to a FM configuration [Fig.~\ref{fig3}(f,g)]. This transition reveals that the central pump alters the coupling between next-nearest neighbor condensates dictating the resulting stable condensate phase configuration. Next, we add two additional barriers at the left- and right-edges of the cluster [Fig.~\ref{fig3}(i)] to demonstrate control of nearest neighbour couplings. The intensity of the side barriers is $\approx30\%$ of the central barrier. Figure~\ref{fig3}(j,k) show the real- and reciprocal-space polariton PL in the presence of all three barriers. From the interference pattern we observe a switching from all FM to a paired ferromagnetic (PFM) coupling. This state is analogous to the compass state reported previously for both scalar~\cite{berloff_realizing_2017, Tan_PRB2018} and spinor polariton condensates~\cite{ohadi_spin_2017}. We note that the transition between different phase ordered states as a function of barrier strength is not digital but instead gradual as a function of barrier strength. Stationary phase configurations, such as presented in Fig.~\ref{fig1} and \ref{fig3}, are separated by non-stationary (cyclical) states (as a function of barrier strength) which can be regarded as a superposition of the aforementioned stable configurations. This is in agreement with recent observations~\cite{topfer_towards_2019}, where regimes of single-mode and multi-mode condensates were found depending on the pumping geometry. In Section~\ref{sec4pumps} we show this transition through numerical simulation between different phases (AFM, FM and PFM) with gradually increasing barrier heights.

\section{Theory}
\begin{figure}[t!]
	\includegraphics[width=1\linewidth]{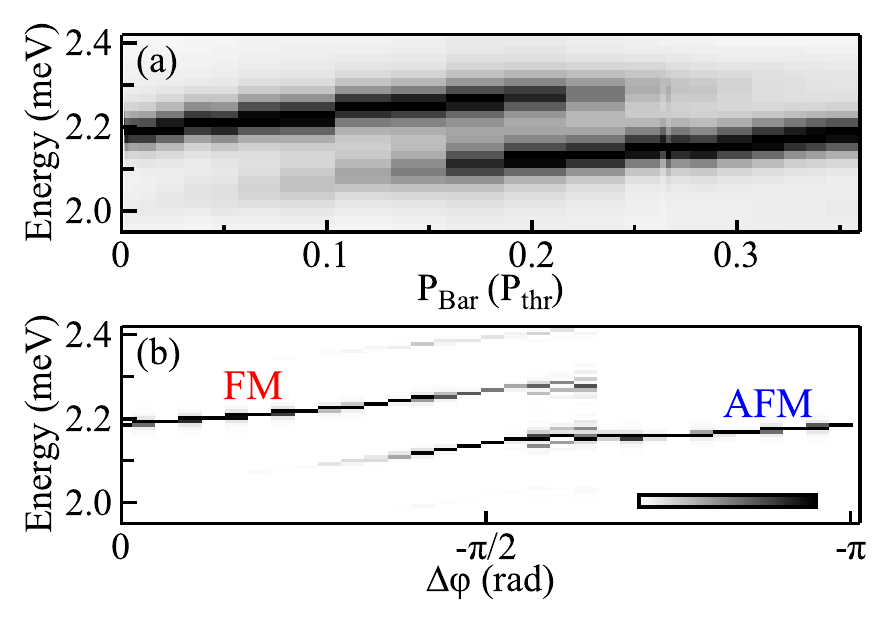}
	\caption{(a) Spectrally resolved PL of the polariton dyad (see Fig.~\ref{fig1}) as a function of applied barrier pump power. (b) Numerically calculated spectra as a function of increasing coupling phase $\beta = \beta^{(0)} + \Delta \phi$ obtained by integrating Eq.\ref{Eq1} with random initial conditions. Parameters: $d = 15.7\;\mathrm{\upmu m}$, $\hbar \Omega = (1.45-i 0.5)\; \mathrm{meV}$, $\hbar \alpha = 0.1\;\mathrm{\upmu eV}$, $\hbar R= 0.5\;\mathrm{\upmu eV}$, $\hbar g= 0.5\;\mathrm{\upmu eV}$, $v = 1.3\;\mathrm{\upmu m}\;\mathrm{ps}^{-1}$, $P=100\;\mathrm{ps}^{-1}$, $\Gamma_R=0.05\;\mathrm{ps}^{-1}$, $\hbar J_0 = 1.1\;\mathrm{meV}$, $k_0=(1.9+i 0.012) \mathrm{\upmu m}^{-1}$ and $\beta^{(0)} = -0.8$ rad.}
	\label{fig4}
\end{figure}

All experimental results are reproduced by numerical integration of the two-dimensional generalized Gross-Pitaevskii equation (see Section~\ref{secGP}). We also reproduce all observations using a recently developed discretized model describing the interacting expanding polariton condensates as time-delay coupled oscillators~\cite{topfer_towards_2019}, i.e.
\begin{align}\notag \label{Eq1}
i \dot{\psi}_n &=  \left[ \Omega  + \left(g + i \frac{R}{2}\right) n_n  + \alpha  |\psi_n|^2  \right] \psi_n  \\
& +  \sum_{m} J_{nm} e^{i\beta_{nm} } \psi_m (t-d_{nm}/v), \\
\dot{n}_n & = -(\Gamma_R + R |\psi_n|^2) n_n + P.
\end{align}
Here $\psi_n$ represents the phase and amplitude of the condensates, $n_n$ their respective reservoirs, and the sum runs over the nearest and next-nearest neighbors. The complex self-energy of each condensate is captured in $\Omega$, the blueshift due to interactions and stimulated scattering rate (i.e., optical gain) from the reservoir is given by $g$ and $R$ respectively, $\alpha$ describes polariton-polariton interaction strength, $\Gamma_R$ the decay rate of the exciton reservoir, $P$ the nonresonant pump power, $J_{nm}$ quantifies the coupling strength between neighboring condensates, $\beta_{nm}$ describes the coupling phase, $v$ is the average phase velocity of polaritons outside their pump spots, and $d_{nm}$ is the distance between neighbors. We use similar parameters as in Ref.~[\onlinecite{topfer_towards_2019}] with the exception of the coupling phase $\beta_{nm}$ which depends on the barrier strength according to Fig.~\ref{fig2}(b-d), i.e. $\beta_{nm} = \beta^{(0)}_{nm} + \Delta \phi$ with phase $\beta^{(0)}_{nm}$ in case of no barrier. For the case of the dyad given in Fig.~\ref{fig1}, the spectrally resolved PL [see Fig.~\ref{fig4}(a)] displays a gradual transition of the dyad mode from a FM to an AFM energy branch as a function of barrier strength, separated by a domain, where both FM and AFM modes are populated. We find that the dominant effect of the barrier is to introduce a phase lag $\Delta \phi$ to the transmitted condensate signal traveling to its nearest neighbor, which ultimately reverses the sign of the complex coupling when $\Delta \phi = -\pi$. Numerically resolving the energies of Eq.~\ref{Eq1} by varying the phase lag $\beta_{nm}$ indeed produces the same behavior in energy as seen in experiment [see Fig.~\ref{fig4}(b)]. The phase lag is mutual in the dyad and we can write for brevity $\beta_{12} = \beta_{21} = \beta$.

\section{Conclusions}
We have developed a new strategy of optically tuning the interactions between polariton condensates in a given two-dimensional network. This approach opens up the path to simulation of synchronization, periodic orbitals, and chaos in more complicated structures with desired nearest neighbour couplings. Moreover, we demonstrated a new experimental technique, which allows single-shot readout of the condensates phases. Implementation of homodyne interferometry is advantageous for phase retrieval in cases where lattice geometry leads to non-trivial phase coupling between adjacent condensates.

{\it Acknowledgements ---} The authors acknowledge the support of the Skoltech NGP Program (Skoltech-MIT joint project), and the UK’s Engineering and Physical Sciences Research Council (grant EP/M025330/1 on Hybrid Polaritonics).

\appendix

\section{Spatiotemporal Simulations} \label{secGP}
\begin{figure}[b!]
	\center
	\includegraphics[width=1\linewidth]{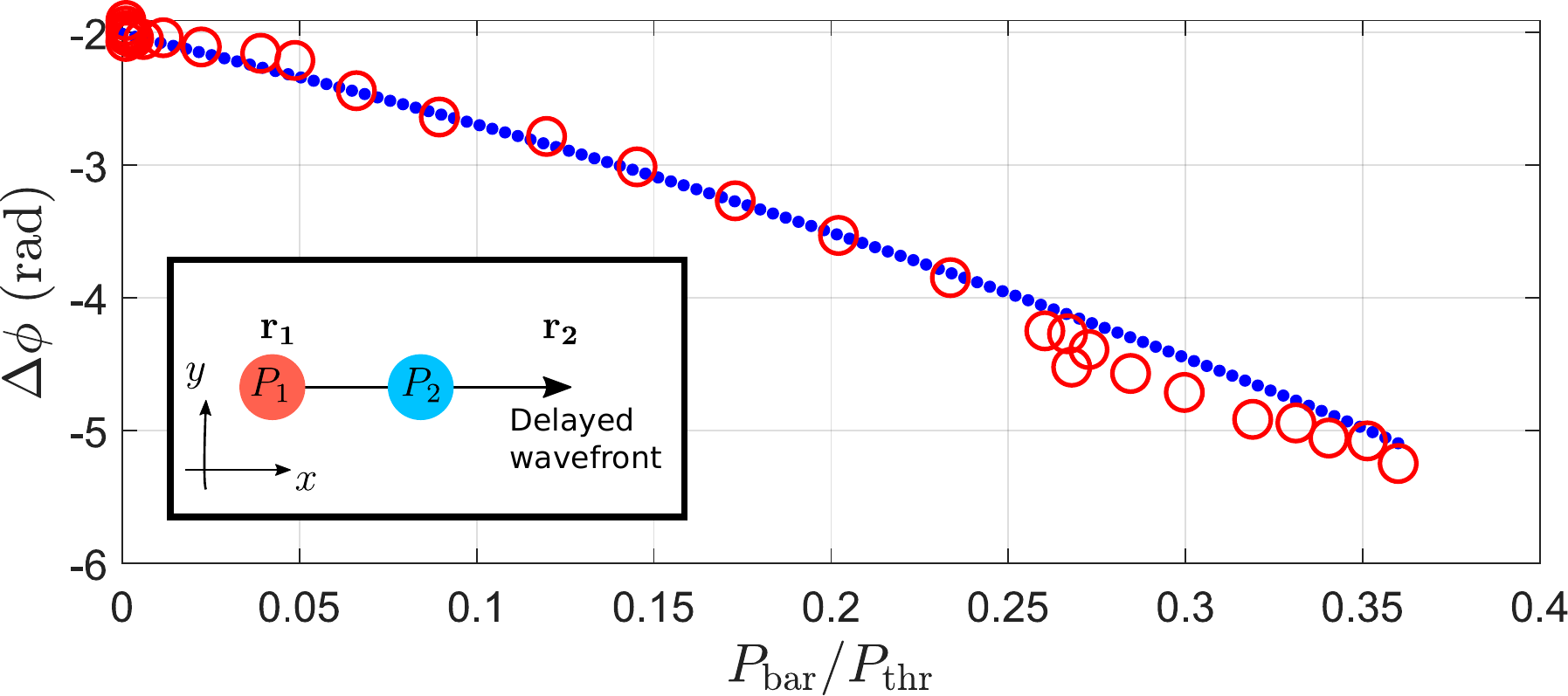}
	\caption{Phase difference between two spatial points of a wavefunction experiencing a barrier (see inset). The wavefunction at the source point $\mathbf{r}_1$ is sustained by a single pump $P = 1.1 P_\text{thr}$ whereas a complex barrier (cross-polarized pump) located in between $\mathbf{r}_1$ and $\mathbf{r}_2$ is gradually increased in power. Results from experiment (red circles) and Eq.~\eqref{eq.GP} (blue dots) show gradual phase accumulation between the two points of the wavefunction.}
	\label{fig1sm}
\end{figure}
The transition from a thermally stochastic state to a macroscopic coherent polariton condensate can be captured within the mean field theory approach. The condensate order parameter $\Psi(\mathbf{r},t)$ is then described by a two-dimensional {\it generalized Gross-Pitaevskii equation},
\begin{equation} \label{eq.GP}
i  \hbar \frac{d \Psi}{d t}  = \left[-\frac{\hbar^2 \nabla^2}{2m} + U(\mathbf{r}) + \alpha |\Psi|^2 - \frac{i \hbar \Gamma }{2} \right] \Psi.
\end{equation}
Here, $m$ is the effective mass of a polariton in the lower dispersion branch, $\alpha$ is the interaction strength of two polaritons in the condensate,  $\Gamma$ is the polariton condensate decay rate, and $U(\mathbf{r})$ is potential energy and gain coming from the pumped reservoir of excitons. The reservoir is conventionally modeled through non-homogeneous rate equations~\cite{Wouters_PRL2007}. We are only interested in stationary states of the interacting condensates and therefore adopt the stationary value of the reservoir written
\begin{align} \notag
U(\mathbf{r}) & =  \left(\frac{1}{2} \frac{g_1 + i R_1}{1 + R_1 |\Psi|^2} + G_1 \right) P(\mathbf{r})\\ &+  \left(\frac{1}{2} \frac{g_2 + i R_2 }{1 + R_2 |\Psi|^2} + G_2 \right) P_\text{bar}(\mathbf{r}).
\end{align}
Where $P$ and $P_\text{bar}$ are the intensities of the co- and cross-polarized pumps respectively. Gain from each pump is then described by the parameters $R_{1,2}$. We note that $P_\text{bar}(\mathbf{r})/(1 + R_2|\Psi|^2) \simeq P_\text{bar}(\mathbf{r})$ since the outflowing condensate $|\Psi|^2$ will be very weak at the positions of the cross-polarized beams. The cross-polarized pump gain $R_2$ can appear as a result of high energy cross-polarized excitons relaxing into a thermal mixture of both spins before feeding into the condensate. Blueshift coming from excitons in the bottleneck regime is described by $g_{1,2}$. This bottleneck blueshift is subject to saturation effects and therefore depends on the condensate in a nonlinear manner. An additional blueshift coming from dark background excitons is described by $G_{1,2}$. 

The polariton mass and lifetime are based on the sample properties: $m = 0.49$ meV ps$^2$ $\upmu$m$^{-2}$, and $\Gamma^{-1} = 6$ ps. We choose values of polariton-polariton interaction strengths typical of GaAs based systems: $\hbar \alpha = 2.4$ $\upmu$eV  $\upmu$m$^2$. The effective reservoir parameters describing blueshift and gain are chosen: $R_1 = 7 \alpha$, $g_1 = 0.8 \alpha$, $G_1 = 4g_1$. The choice of these parameters is somewhat ambiguous given that the fraction of blueshift coming from dark inactive excitons against radiative active excitons is not clear to date. The above choices are made to fit the standard experimental observation of a sharp ring in $k$-space around $k \approx 1.9$ $\upmu$m$^{-1}$ at threshold, and $\sim 350$ $\upmu$eV increase in energy when raising the pump power to $1.5 \times$ threshold power. The value of the cross-polarized scattering rate is set to $R_2 = 0.3 R_1$ similar to the value used in experiments on horizontally nonresonantly driven polariton condensates~\cite{Redondo_NJP2018}. The pumps are Gaussian shaped but differ in width, which is denoted by their RMS value $w_{1,2}$ where $w_1 = 1.06$ $\upmu$m (corresponding to $\approx 2.5$ $\upmu$m FWHM) and $w_2 = 1.6 w_1$. The pump spots are chosen slightly larger than in experiment to account for the finite diffusion of the reservoir away from the laser beam.

\section{Phase lag through a barrier} \label{sec1pump}
\begin{figure}[h!]
	\center
	\includegraphics[width=1\linewidth]{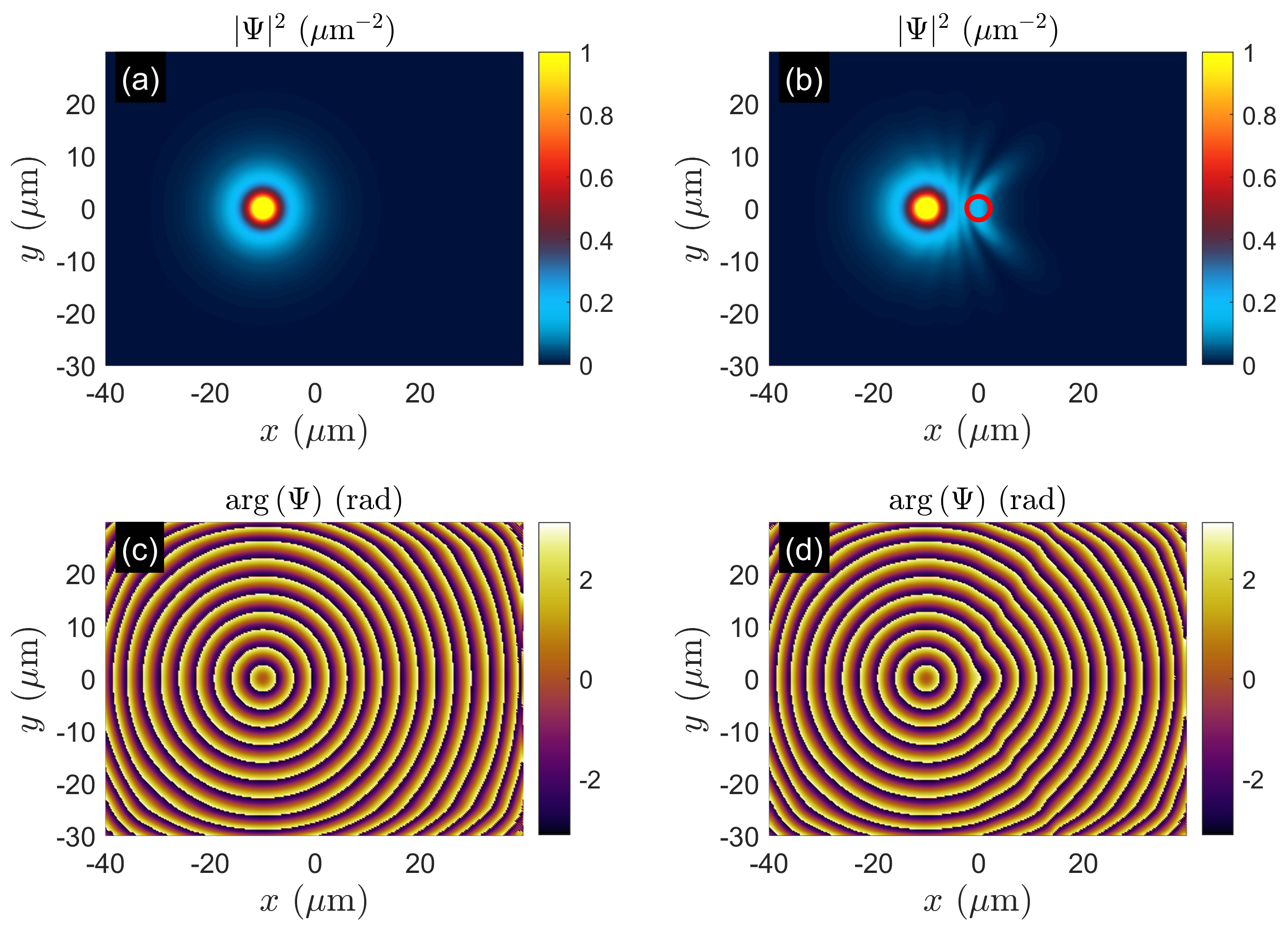}
	\caption{(a,b) Density and (c,d) phase maps of the condensate wavefunction before (a,c) and after (b,d) activation of the barrier at $P_\text{bar}/P_\text{thr} = 0.36$ power. One can notice strong modulations in the density (shown in a saturated linear colorscale) corresponding to the wavefront scattering of the barrier and a phase lag appearing. Red circle in panel (b) denotes the location of the barrier $P_\text{bar}$. Panels (a,b) are normalized and shown in a linear colorscale.}
	\label{fig2sm}
\end{figure}

Here we consider the same scenario as given in Fig.~\ref{fig2}(b-d) where a single condensate is driven above threshold with a pump $P$ and a barrier is introduced with a cross-polarized pump $P_\text{bar}$ (inset in Fig.~\ref{fig1sm}). We are interested in the phase accumulation $\Delta \phi = \arg{(\Psi(\mathbf{r}_2) \Psi^*(\mathbf{r}_1))}$ when the condensate wavefunction expands from its source point to its mirror point about the barrier. The results are given in Fig.~\ref{fig1sm} which show a gradual decrease in phase accumulation corresponding to phase lag caused by the increasing pump barrier. We point out that the results are extracted after the wavefunction has converged to its steady state. In Fig.~\ref{fig2sm} we show spatial colormaps of the wavefunction density (a,b) and phase (c,d) for $P_\text{bar}/P_\text{thr} = 0$ and $0.36$ respectively. In what follows, by setting $g_2 = 1.8 \alpha$ and $G_2 = 4 g_2$ we reproduce all observations of the experiment.
\begin{figure}[t!]
	\center
	\includegraphics[width=1\linewidth]{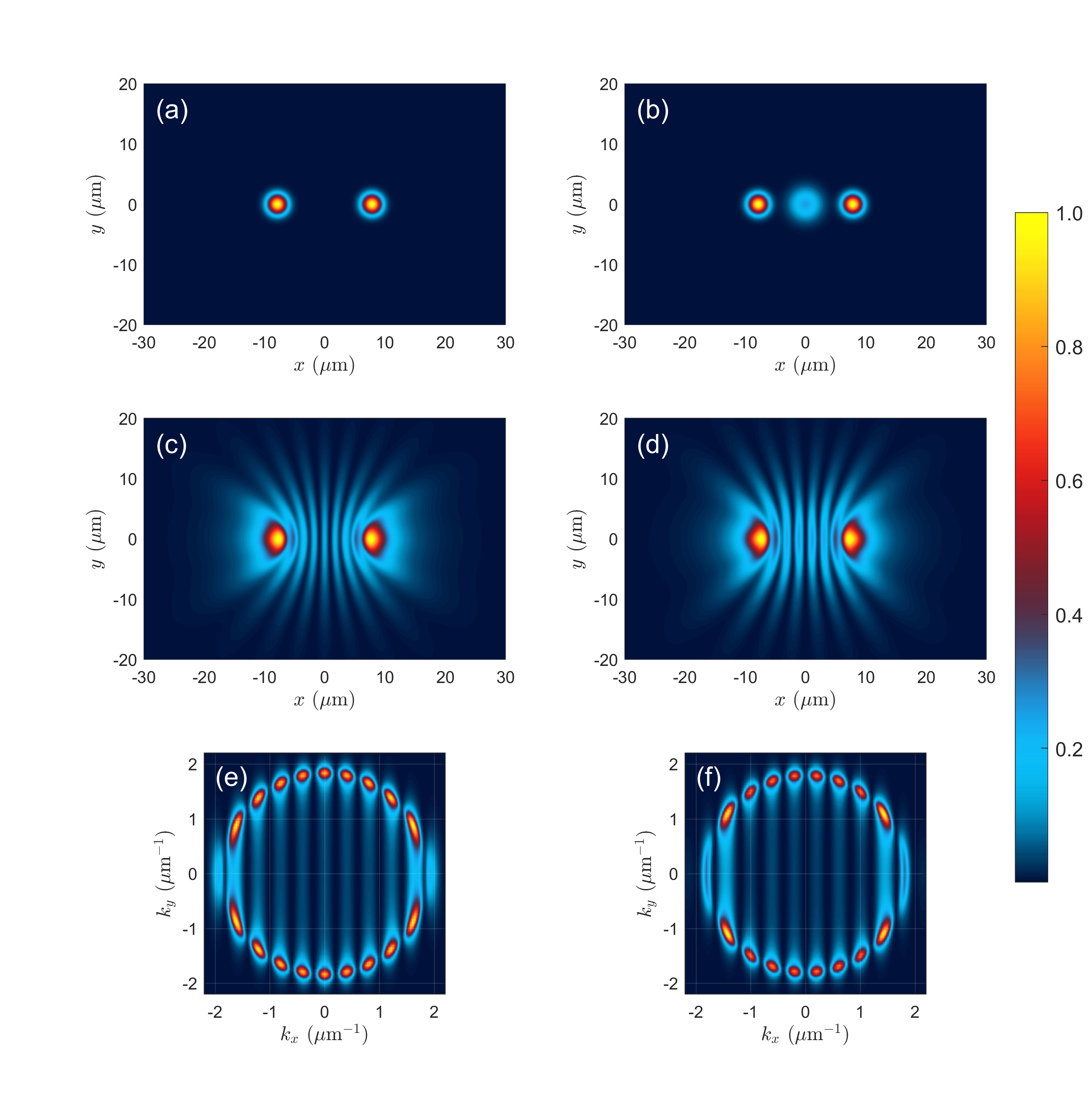}
	\caption{(a,b) Pump profiles, (c,d) condensate real space density and (e,f) reciprocal space density. In (a,c,e) the barrier is switched off whereas in (b,d,f) the barrier is set to $P_\text{bar}/P_\text{thr} = 0.25$ power. The barrier introduces modified interference to the system which favors condensation into a lower energy mode. Data in all panels is normalized and in a linear colorscale.}
	\label{fig3sm}
\end{figure}
\begin{figure}[b!]
	\center
	\includegraphics[width=1\linewidth]{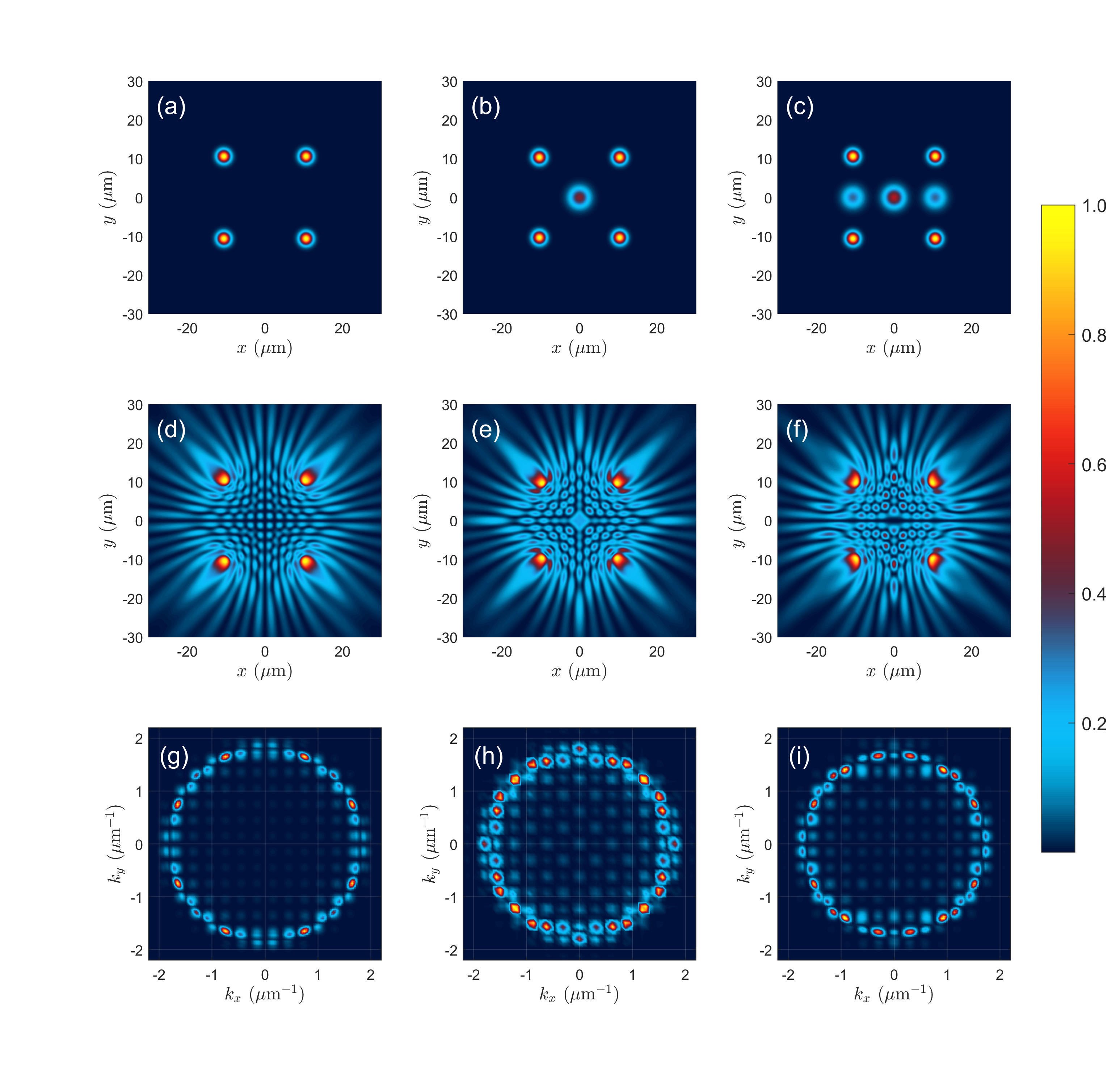}
	\caption{(a,b,c) Pump profiles, (d,e,f) condensate real space density and (g,h,i) reciprocal space density. In (a,d,g) the barrier is switched off whereas in (b,e,h) the central barrier is set to $P_\text{bar}/P_\text{thr} = 0.53$ power, and in (c,f,h) the central barrier remains at $P_\text{bar}/P_\text{thr} = 0.53$ but with side barriers at $P_\text{bar}/P_\text{thr} = 0.32$. The barriers introduce modified interference to the system which favors condensation into the lower energy modes. Data in all panels is normalized and in a linear colorscale.}
	\label{fig4sm}
\end{figure}

\section{Two condensates with a barrier} \label{sec2pumps}
In Fig.~\ref{fig3sm} we reproduce the results of Fig.~\ref{fig1} by direct numerical integration of Eq.~\eqref{eq.GP} from a stochastic (white noise) initial condition. Panels (a,b) show the pump profile $P + P_\text{bar}$ without and with $P_\text{bar} \neq 0$ respectively. Panels (c,e) show the real- ($|\Psi(\mathbf{r})|^2$) and reciprocal ($|\hat{\Psi}(\mathbf{k})|^2$) density of the steady state wavefunction when no barrier is present, resulting in a formation of a FM state. When the barrier is switched on we see in panels (d,e) that the new steady state belongs to an AFM configuration.  

\section{Four condensates with three barriers} \label{sec4pumps}
In Fig.~\ref{fig4sm} we reproduce the results of Fig.~\ref{fig3} by direct numerical integration of Eq.~\ref{eq.GP} from a stochastic (white noise) initial condition. Panels (a,b,c) show the total pump profile $P + P_\text{bar}$ of co- and cross-polarized beams. Panels (d,g) show the real- ($|\Psi(\mathbf{r})|^2$) and reciprocal ($|\hat{\Psi}(\mathbf{k})|^2$) density of the steady state wavefunction when no barrier is present, resulting in a formation of a AFM state across vertical and horizontal axes. When a central barrier is switched on we see in panels (e,h) that the new steady state belongs to an FM configuration across vertical and horizontal axes. When additional barriers are activated on the left and right side of the $2\times2$ condensate cluster we obtain the PFM state.

\bibliography{Library}
\end{document}